\documentclass[a4paper,11pt]{article}
\pdfoutput=1 

\usepackage{jcappub} 

\usepackage[english]{babel}
\usepackage[utf8]{inputenc}
\usepackage{float}
\usepackage{graphicx}
\usepackage[caption=false]{subfig}
\usepackage{color}
\usepackage{upgreek}

\usepackage{amsfonts}
\usepackage{amsmath}
\usepackage{amssymb}
\usepackage{graphicx}
\usepackage{mathrsfs}

\usepackage{ulem}

\title{\boldmath Unveiling the nuclear matter EoS from neutron star properties: a supervised machine learning approach}

\author[a]{M\'arcio Ferreira,}
\author[a]{Constan\c ca Provid\^encia}

\affiliation[a]{CFisUC, Department of Physics, University of Coimbra, P-3004 - 516  Coimbra, Portugal}

\emailAdd{marcio.ferreira@uc.pt}
\emailAdd{cp@fis.uc.pt}

\abstract{We explore supervised machine learning methods in extracting the non-linear maps between neutron stars (NS) observables and the equation of state (EoS) of nuclear matter.	Using a Taylor expansion around saturation density, we have generated a set of model independent EoS describing stellar matter constrained by nuclear matter parameters that are thermodynamically consistent, causal, and consistent with astrophysical observations. From this set, the full non-linear dependencies of the NS tidal deformability and radius on the nuclear matter parameters were learned using two distinct machine learning methods. Due to the high accuracy of the learned non-linear maps, we were able to analyze the impact of each nuclear matter parameter on the NS observables, identify dependencies on the EoS properties beyond linear correlations and predict which stars allow us to draw strong constraints.}

\begin{document}
\maketitle
\flushbottom

\section{Introduction}
The properties of the equation of state (EoS) of nuclear matter at
supra-saturation densities still remain an open question in nuclear
physics. Neutron stars (NSs) become unique astrophysical objects
through which the properties of super-dense neutron-rich nuclear
matters can be studied. Constraining the EoS is a combined task
between astrophysics and nuclear physics. Astrophysical observables
become important probes for the dense nuclear matter
properties. In particular, two-solar mass NS detected during the last
ten years set quite stringent constraints on EoS of nuclear matter. 
The pulsar PSR J1614-2230 is, among the most massive observed pulsars, 
the one with the smallest uncertainty on the mass $M=1.908\pm0.016\, M_\odot$ \cite{Arzoumanian2017,Fonseca2017,Demorest2010}. 
Other two pulsars with a mass above  two solar masses are PSR J0348+0432 with $M=2.01 \pm 0.04 M_\odot$ \cite{Antoniadis2013} 
and the recently detected   MSP J0740+6620 with a mass $2.14^{+0.10}_{-0.09}M_\odot$ \cite{Cromartie2019}, 
both masses given within a 68.3\% credibility interval. \\

The coalescence of binary NS systems is an important source of gravitational waves (GW) \cite{Cutler1993,Cutler2002,Radice2016,TheLIGOScientific:2017qsa,Abbott:2018exr}. 
The energy emitted in GW crucially depends on the EoS of NS matter, allowing to constrain the 
nuclear matter EoS from GW observations. One crucial information carried by GW is the tidal 
deformability of the NS \cite{Flanagan2008,Hinderer2007}. During the last inspiral stage of a 
coalescing binary NS system, each NS develops a mass-quadrupole moment due to the extremely strong 
tidal gravitational field of the companion NS. The ratio of the induced quadrupole moment to the 
external tidal field is proportional to the tidal deformability $\Lambda$. 
It is an important observable that is sensitive to the nature of the EoS.\\

The inspiral compact binary event GW170817 was the first binary NS inspiral observed by the LIGO--Virgo collaboration \cite{TheLIGOScientific:2017qsa}. Due to its potential to directly probe the physics of NSs, 
it was acknowledged as the beginning of a new era in the field of multi-messenger astronomy and nuclear physics.
The analysis of GW170817 has placed upper bounds on the NSs combined dimensionless tidal deformability. 
Using a low-spin prior (consistent with the observed neutron star population), a combined dimensionless 
tidal deformability of the NS merger was obtained,  $\tilde{\Lambda}\le800$ with 90 \% confidence.
A reanalysis of GW170817 was made assuming the same EoS and spins within 
the range observed in Galactic binary neutron stars, in which the tidal deformability of a $1.4 M_{\odot}$ NS was estimated to be $70<\Lambda_{1.4M_{\odot}}<580$ at the 90\% level \cite{Abbott:2018exr}.\\

Correlation analyses between nuclear matter parameters and NS properties have been explored using several nuclear models \cite{Vidana2009,Xu2009,Ducoin2010,Ducoin2011,Newton2013,Alam2016,Malik:2018zcf,Carson:2018xri,Zhang:2018vrx,Zimmerman2020,Li2020}.
These correlation studies show, however, a considerable model dependence and are only sensitive to linear dependencies. 
One possible model-independent way of accessing the properties of the EoS is by parameterizing the EoS of asymmetric nuclear matter 
  around the saturation density \cite{Margueron2017,Carson:2018xri,Zhang:2018vrx}. 
The analysis of the impact of each empirical nuclear matter parameter on NS properties is an intricate problem due to the 
dimensionality of the input space and the interdependencies among the parameters. One can proceed by reducing the input 
space by fixing some parameters \cite{Zhang_2018, Raithel_2019}. However, due to the nuclear matter parameter uncertainties, 
the results obtained from a dimensional reduction of the input space depend on the selected parameter values.
Some specific non-linear dependencies of NS properties on nuclear matter parameters were studied in \cite{Sotani2013,Silva2016,Carson:2019xxz}.\\

In the present work, we explore supervised a Machine Learning (ML) framework to learn the non-linear maps between 
the EoS and NS observables (we focus on the NS tidal deformability and radius). 
These methods allow us to search for all kinds of dependencies and interactions among nuclear matter parameters 
that explain NS properties regardless of the input space dimensionality.
Unlike correlation analyses, these non-linear maps, which encode the full dependency of NS observables 
on the EoS, enable us to measure the effect of each empirical parameter.

The supervised ML methodology has already been applied to NS physics in \cite{Fujimoto:2017cdo}, where a Deep Neural Network (DNN) was used as an efficient procedure for mapping from a finite set of mass-radius observational data onto the EoS plane. The analysis was later extended \cite{Fujimoto:2019hxv} by comparing the DNN predictions with the conventional approaches for the nuclear EoS and on the tidal deformability bound. In \cite{Morawski:2020izm}, the authors analyzed an auto-encoder architecture for reconstructing the neutron star EoS using observable parameters: masses, radii, and tidal deformabilities. A different approach has been recently proposed to constrain the EoS from observations through the introduction of a non-parametric EoS representation \cite{Landry2018,Essick2019,Landry2020,Essick2021}.\\

In the following, we restrict our study to nucleonic EoS that may be described by a continuous function. Observations that cannot be reproduced by our set of EoS may indicate that a more complex EoS, including phase transitions, describes matter inside the compact star.
\\

The paper is organized as follows. In Sec. 2, we briefly
describe the two supervised ML methods used in the present work. In Sec. 3, we outline the EoS parametrization employed.
The EoS generating process and the final EoS dataset, used in the supervised learning, is summarized in Sec. 5.  
The learning procedure is described in Sec. 5 and the results are shown in Sec. 6. Finally the conclusions are drawn in Sec. 7.

\section{Supervised Machine Learning}
Our goal is to explore ML methods to learn the non-linear maps between the EoS of the nuclear matter and astrophysical observables. We will apply two supervised ML methods to access these non-linear maps. 
Below, we give a very general and brief introduction to these methods (for a rigorous exposition see \cite{friedman2001elements,scholkopf2001learning,goodfellow2016deep}).   
\subsection{Support Vector Machines Regression}
Support Vector Machines (SVM) is a well-known supervised ML  method for classification problems. It is a maximum-margin classifier: it searches for the hyperplane (decision boundary) that separates the classes and has the maximum possible margin. 
The SVM are easily extended to non-linear classifiers using kernel functions. They are non-linear maps that transform a non-linearly separable input space into a higher dimensional space where a linear separation is possible. It might be impossible, however, to find a hyperplane that perfectly separates the classes using only the input features. Then, a further generalization that allows margin violations is introduced using slack variables (soft-margin classifiers). A balance between having the largest possible margin while limiting the margin violations is searched.
Support Vector Machines Regression (SVM-R) is the application of the above maximum-margin idea of SVM to regression problems. 
SVM-R searches for the maximum margin (hyper-tube of width $\epsilon$) that encloses the maximum number of points while limiting margin violations (for a $\epsilon$-insensitive loss regression). The regression model is the hyperplane positioned right at the center of this hyper-tube.\\

\subsection{Deep Neural Networks}
DNN are widely-used nonlinear models for supervised learning. A DNN consists of hierarchical layers made of neurons (the basic unit). In Feed-forward DNN, the input vector enters the first layer (input layer) and then proceeds sequentially through the middle layers (hidden layers) up to the last layer (output layer). At each neuron, a vector of $d$ features, $\textbf{x}=(x_1,...,x_d)$, is transformed into a scalar output $y$ by two basic transformations: a linear transformation, which weighs the relative importance of its inputs, and a non-linear transformation (activation function). First, the linear transformation $z=\textbf{w}.\textbf{x}+b$ is computed, where $\textbf{w}=(w_1,w_2,...,w_d)$ are the neuron weights (it receives $d$ input features), and $b$ is the neuron bias. Then, the linear transformation $z$ passes through the activation function $\sigma(z)$. Some common choices are the rectified linear function, $\sigma(x)=\max\{0,x\}$, and the hyperbolic tangent, $\sigma(x)=\tanh(x)$. The output of one layer enters as input in the next layer. This procedure is repeated until the output layer is reached. The output of a DNN is then 
a complex non-linear transformation of the inputs that depends on all neuron weights and biases that compose all layers. Training a DNN consists in minimizing a loss function by the gradient descent method in order to find the optimal weights and biases.
Herein, we use as loss function the Mean Squared Error (MSE), $L(\boldsymbol{w,b})=(1/N)\sum_{i}^N(\widehat{y}_i(\boldsymbol{w,b})-y_i)^2$, where $y_i$ is the actual output, $\widehat{y}_i(\boldsymbol{w,b})$ is the DNN's prediction, and $N$ is the number of points considered at each step of the minimization procedure.\\

\section{Equations of state}

We parametrize the equation of state (EoS) of nuclear matter from a Taylor
expansion around the saturation density. This has been frequently done
in the past, see for instance
\cite{Furnstahl2001,Baran2004,Vidana2009,Ducoin2011},  and also recently
\cite{Margueron2017,Zhang:2018vrx,Li2020}.   The coefficients of the expansion
are identified with  empirical
parameters that characterize nuclear matter.
Expressing the EoS  in this form will allow
us to analyze the properties of neutron stars systematically  by
varying the empirical parameters 
continuously within  their uncertainty range. In the present study we
will consider the expansion up to  third order on the density.\\

We start from the generic functional form for the energy per particle of homogeneous nuclear matter 

\begin{equation}
{\cal E}(x,\delta)=e_{\text{sat}}(x)+e_{\text{sym}}(x)\delta^2,
\end{equation}
and perform a Taylor expansion on both $e_{\text{sat}}(x)$ and $e_{\text{sym}}(x)$ up to third order,
\begin{align}
e_{\text{sat}}(x)&=E_{\text{sat}}+\frac{1}{2}K_{\text{sat}}x^2+\frac{1}{6}Q_{\text{sat}}x^3\\
e_{\text{sym}}(x)&=E_{\text{sym}}+L_{\text{sym}}x+\frac{1}{2}K_{\text{sym}}x^2+\frac{1}{6}Q_{\text{sym}}x^3
\end{align}
where $n=n_n+n_p$ is the baryonic density, $\delta=(n_n-n_p)/n$ is the asymmetry, and 
$x=(n-n_{0})/(3n_{0})$. The neutron and proton densities are denoted by $n_n$ and $n_p$, respectively.
The isoscalar empirical parameters are defined as successive density derivatives of $e_{\text{sat}}(x)$,
\begin{equation}
P_{\text{IS}}^{(k)}=(3n_{0})^k\left.\frac{\partial^k e_{\text{sat}}}{\partial n^k}\right|_{\{\delta=0,n=n_{0}\}},
\end{equation}
whereas the isovector parameters measure density derivatives of $e_{\text{sym}}(x)$,
\begin{equation}
P_{\text{IV}}^{(k)}=(3n_{0})^k\left.\frac{\partial^k e_{\text{sym}}}{\partial n^k}\right|_{\{\delta=0,n=n_{0}\}}.
\end{equation}
The corresponding empirical parameters are then 
\begin{equation}
\{E_{\text{sat}},K_{\text{sat}},  Q_{\text{sat}}\}  \rightarrow
\{P_{\text{IS}}^{(0)},P_{\text{IS}}^{(2)},P_{\text{IS}}^{(3)}\}
\end{equation}
and
\begin{equation}
\{E_{\text{sym}},L_{\text{sym}}, K_{\text{sym}},  Q_{\text{sym}}\}  \rightarrow
\{P_{\text{IV}}^{(0)},P_{\text{IV}}^{(1)},P_{\text{IV}}^{(2)},P_{\text{IV}}^{(3)}\}.
\end{equation}
Being a rather well determined quantity and to reduce the computational cost, we have fixed the value of the saturation energy $E_{\text{sat}}=-15.8$ MeV (the current estimated value is  $-15.8\pm0.3$ MeV \cite{Margueron2017}). Furthermore, the saturation density is also fixed at $n_{0}=0.155$ fm$^{-3}$. Therefore, 
each EoS is represented by a point in a 6-dimensional input space,
$$\text{EoS}_i \rightarrow (K_{\text{sat}},  Q_{\text{sat}}, E_{\text{sym}},L_{\text{sym}}, K_{\text{sym}},  Q_{\text{sym}}).$$ 

The Taylor expansions, Eqs (1)-(3), are widely used to characterize nuclear matter around saturation density in experimental analysis.
The coefficients are useful to extract local information of different nuclear models around the saturation density. 
The expansion is only accurate around saturation density and it becomes increasingly unreliable for large densities. 
We consider these Taylor expansions, however, as simple parametrizations, similar to many other EoS parameterizations, with the characteristic of becoming Taylor expansions when $n \rightarrow n_0$ \cite{Zhang2018}.  The parameters are just unknowns and there is thus no convergence issue. The advantage of the present parametrization is that
it allows to constrain the EoS directly from the properties of symmetric nuclear matter and the symmetry energy near saturation density, while other parametrizations indirectly impose such properties \cite{Zhang2018}. 
On the other hand, the higher-order parameters obtained by applying the present parametrization to dense matter may substantially deviate from the actual nuclear matter expansion coefficients, and should be seen as effective parameters
incorporating the effect that the higher order missing terms should reproduce. In \cite{Margueron2017},   it was shown that a known EoS obtained from  either a Skyrme force or within the RMF framework, could be well reproduced taking terms in the Taylor expansion until third or forth order.\\

In a recent work \cite{Ferreira2020}, we have studied the effect of
matching the crust to the core EoS using three different crust EoS together with a forth order Taylor expansion. It was shown that there are  uncertainties intrinsically connected to the matching of the
crust to the core EoS when non-unified EoS are used. The average nuclear matter properties and standard
deviations associated with the EoS that satisfy the same constraints
we use in the present study depend only slightly on the crust. We have verified that these properties are very similar to the ones obtained with a third order Taylor expansion, and the only quantity that may deviate being $Q_{\text{sat}}$.  We, therefore, consider that the six independent parameters introduced give  the desirable freedom to study high density matter of a nucleonic star, and do not require the computational effort of a eight parameter EoS. However, taking into account the possible limitations of the present approach to describe high density matter, we will restrict the following discussion to low mass NS properties.\\

The probability of an EoS is given by a multivariate Gaussian distribution with diagonal covariance matrix.
We do not impose correlations on the empirical parameters a priori (the off-diagonal elements of the covariance matrix vanish). The physical correlations among the empirical parameters arise from a set of physical constraints  \cite{Margueron2017,Margueron:2017lup}.
A valid EoS must fulfil the following conditions: i) be monotonically increasing (thermodynamic stability);
ii) the speed of sound must not exceed the speed of light (causality);
iii)  supports a maximum mass at least as high as $1.97M_{\odot}$ \cite{Arzoumanian2017,Fonseca2017,Demorest2010,Antoniadis2013} (observational consistency);
iv) predicts a tidal deformability of $70<\Lambda_{1.4M_{\odot}}<580$ \cite{Abbott:2018exr} (observational consistency);
and v) the symmetry energy $e_{\text{sym}}(x)$ is positive.
We also analyze the impact of imposing $M\geq 2.05M_\odot$ \cite{Cromartie2019}.
All the EoS are in $\beta$-equilibrium.
For the low density region we use the SLy4 EoS \cite{Douchin2001}. To be considered a valid EoS, the generated EoS must cross the 
SLy4 EoS in the $P(\mu_B)$ plane at densities $n<n_{0}=0.155$ fm$^{-3}$, where $\mu_B$ is the baryon chemical potential. 
The transition density between the crust 
and the hadronic EoS of the final set turns out to be strongly clustered around $n\approx 0.1$ fm$^{-3}\sim2n_0/3$.
The crust-core transition occurs, depending on the model, for densities in the
range $1/3\lesssim n/n_0 \lesssim 2/3$ \cite{Ducoin2011,Newton2013}. We have allowed for a slightly larger transition density, in particular $n_0$,  because: (i) the crust and
core are not defined in a unified way and, therefore, it is expected
that the matching may occur out of the expected crust-core
transition density range; (ii) in
order to get information on the nuclear matter properties at
saturation from astrophysical observations the effect of the crust
EoS should be constrained to densities below saturation.
We calculate both the mass-radius relation,
by solving the Tolman-Oppenheimer-Volkoff (TOV) \cite{TOV1,TOV2} and the tidal deformability $\Lambda$ \cite{Hinderer2007} equations.

\section{EoS Dataset}
We have generated $10^8$ EoS by sampling from the 6-dimensional EoS parameter space
\begin{equation}
\text{EoS} = (K_{\text{sat}}, Q_{\text{sat}},E_{\text{sym}},L_{\text{sym}}, K_{\text{sym}},  Q_{\text{sym}} ) = (X_1, \dots, X_6)  \sim N({\boldsymbol X};\boldsymbol{\mu},\boldsymbol{\Sigma}),
\end{equation}
    where $\boldsymbol{\mu}$ is the mean vector with
	components $\mu_i=\text{E} [X_i]$ 
	\footnote{$\text{E}$ is the expected value operator, $\operatorname {E} [X_i]=\int X_i p(X_i)\,dX_i$, where $p(X_i)=N(X_i;\mu,\Sigma)$ is the marginal probability of $X_i$} and $\boldsymbol{\Sigma}$ is the covariance matrix with entries $\Sigma_{ij}=\text{E} [(X_{i}-\mu _{i})(X_{j}-\mu _{j})]$, which characterize the multivariate Gaussian distribution $N({\boldsymbol X};\boldsymbol{\mu},\boldsymbol{\Sigma})$. We do not consider correlation between the empirical parameters  a priori, i.e., $\Sigma_{ij}=0$ for $i \neq j$. The values for $\mu_i$ and $\Sigma_{ij}$ are given in Table \ref{tab1:gaussian_para}.
The standard deviation values $\sqrt{\Sigma_{ii}}$  reflect a global estimation for the empirical parameters \cite{Margueron2017}.

\begin{table}[!htb]
	\begin{center}
		\begin{tabular}{ccc}
			\hline
			& $\mu_i$ [MeV] & $\sqrt{\Sigma_{ii}}$ [MeV]\\
			\hline
			\hline
			$K_{\text{sat}}$ & 230 & 20  \\
			$Q_{\text{sat}}$ & 300 & 400  \\
			$E_{\text{sym}}$ & 32 & 2    \\
			$L_{\text{sym}}$ & 60 & 15   \\
			$K_{\text{sym}}$ & -100 & 100  \\
			$Q_{\text{sym}}$ & 0 & 400  \\
			\hline
		\end{tabular}
		\caption{The mean $\mu_i$ and standard deviation $\sqrt{\Sigma_{ii}}$ of the multivariate Gaussian, where $\Sigma_{ii}$ are the variances, i.e., the diagonal elements of the covariance matrix $\boldsymbol{\Sigma}$. The off-diagonal elements $\Sigma_{ij}$ ($i \neq j$) are zero. Values taken from \cite{Margueron2017}.}
		\label{tab1:gaussian_para}
	\end{center}
\end{table}

After all the constraints have been applied to each sampled EoS, we ended up with a set of 13038 EoS. The sample mean and standard deviation for the empirical parameters for this final set are given in Table \ref{tab2:statistics}.

\begin{figure}[!htb]
	\centering
	\includegraphics[scale=0.4]{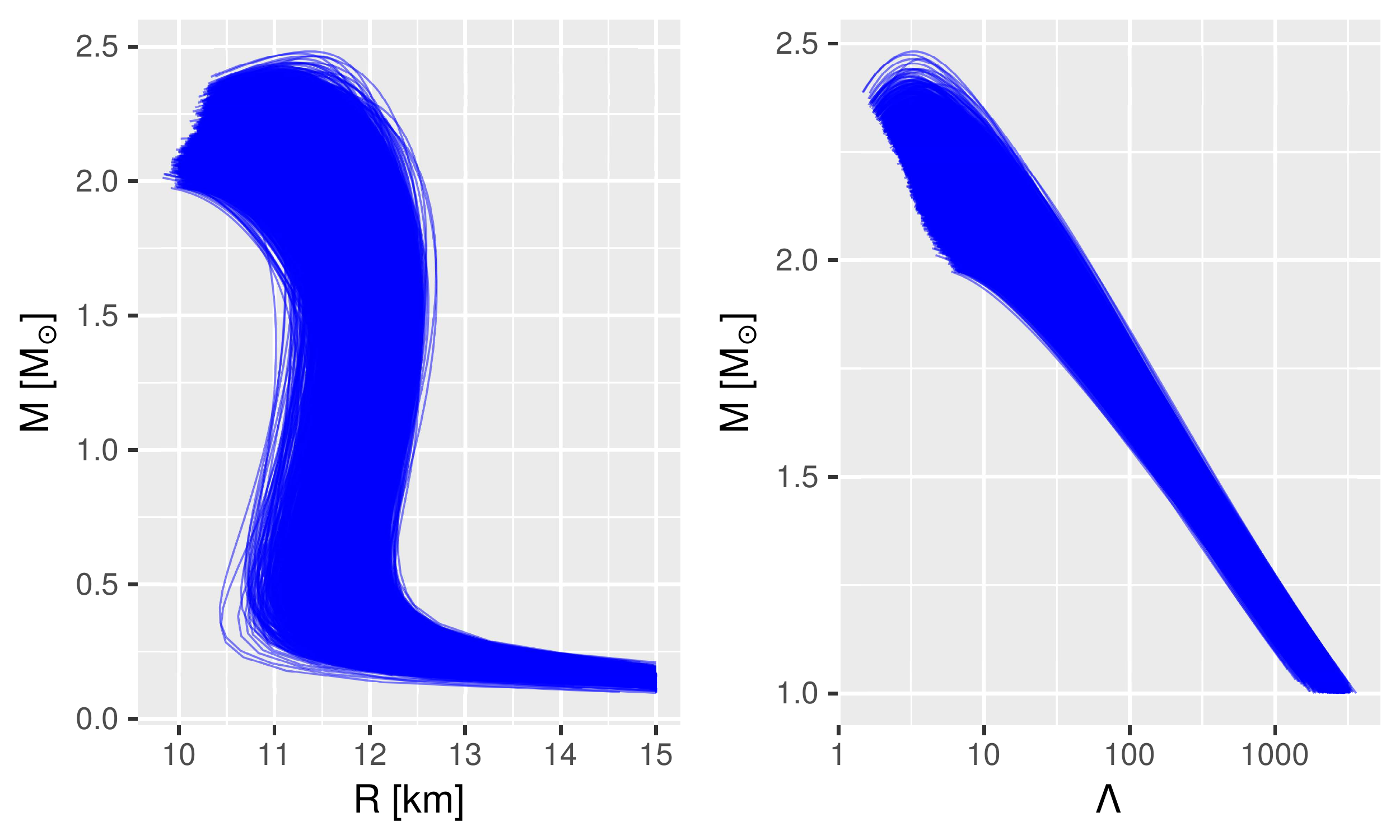} 
	\caption{Mass-radius (left) and mass-$\Lambda$ (right) relations for the set of $N=13038$ EoS.}
	\label{fig1:m-r}
\end{figure}

\begin{table}[!htb]
	\centering
	\begin{tabular}{rrr}
		\hline
		& $\mu_i$ [MeV] & $\sqrt{\Sigma_{ii}}$ [MeV]\\
		\hline
		\hline
		$K_{\text{sat}}$  & 232.92 & 18.37 \\ 
		$Q_{\text{sat}}$  & -94.02 & 29.57 \\ 
		$E_{\text{sym}}$  & 33.20 & 1.83 \\ 
		$L_{\text{sym}}$  & 51.85 & 11.25 \\ 
		$K_{\text{sym}}$  & -60.83 & 55.74 \\ 
		$Q_{\text{sym}}$  & 290.03 & 215.96 \\ 
		\hline
	\end{tabular}
	\caption{Sample mean $\mu_i=(1/N)\sum_i x_i$ and 
		standard deviation $\sqrt{\Sigma_{ii}}=\sqrt{\sum_i(x_i-\bar{x})^2/(N-1)}$ for the empirical parameters of the $N=13038$ valid EoS.}
	\label{tab2:statistics}
\end{table}
The obtained $M-R$ and $M-\Lambda$ diagrams, for the entire set of 13038 EoS, are shown in Fig. \ref{fig1:m-r}. 
The mean values of $\Lambda_{M_i}$ and $R_{M_i}$ and their standard deviations for $1.0M_{\odot}$ and $1.4M_{\odot}$ are given in Table \ref{tab3:statistics_lambda_R}.

\begin{table}[!htb]
	\centering
	\begin{tabular}{rrr}
		\hline
		& $\mu_i$  & $\sqrt{\Sigma_{ii}}$ \\
		\hline
		$\Lambda_{1.0 M_{\odot}}$ & 2808.887 & 261.523 \\ 
		$\Lambda_{1.4 M_{\odot}}$ & 415.723 & 46.179 \\ 
		\hline
		& $\mu_i$ [km]  & $\sqrt{\Sigma_{ii}}$ [km] \\
		\hline
		$R_{1.0 M_{\odot}}$ & 11.950 & 0.190 \\ 
		$R_{1.4 M_{\odot}}$ & 12.055 & 0.194 \\ 
		\hline
	\end{tabular}
	\caption{Sample statistics (mean and standard deviation values) for $\Lambda_{M_i}$ and $R_{M_i}$ from the dataset of 13038 EoS.}
	\label{tab3:statistics_lambda_R}
\end{table}

The resulting distribution for the NS radius is quite narrow,
e.g., for the canonical mass of $1.4M_{\odot}$, the dispersion is only $0.19$ km around the mean value $12.06$ km (see Table \ref{tab3:statistics_lambda_R}). 
For the same NS mass, the tidal deformability is centered around 415.72  with a dispersion of $46.18$. To summarize, the final dataset contains $13038$ EoS (points) in a 6-dimensional input space, $\{K_{\text{sat}}, Q_{\text{sat}},E_{\text{sym}},L_{\text{sym}},K_{\text{sym}}, Q_{\text{sym}}\}$, and the corresponding response variables (NS properties), $\Lambda_{M_i}$ and $R_{M_i}$, we want to explore in the next sections.

\section{Learning Procedure}
Our goal is to use DNN and SVM-R models to learn the non-linear maps between the EoS, i.e., the empirical parameters of nuclear matter,
and neutron star observables. Herein, we focus on the neutron star radius and tidal deformability as a function of the neutron star mass.
Therefore, the non-linear maps one wants to learn are:
\begin{align}
\widehat{\Lambda}_{M_i}&(K_{\text{sat}}, Q_{\text{sat}},E_{\text{sym}},L_{\text{sym}},K_{\text{sym}}, Q_{\text{sym}}), \label{lmi}\\
\widehat{R}_{M_i}&(K_{\text{sat}}, Q_{\text{sat}},E_{\text{sym}},L_{\text{sym}},K_{\text{sym}}, Q_{\text{sym}}). \label{rmi}
\end{align}
We use the symbol $\,\,\,\widehat{}\,\,\,$ to distinguish the model's predictions,  $\widehat{\Lambda}_{M_i}$ and $\widehat{R}_{M_i}$, from the {\it true} values, $\Lambda_{M_i}$ and $R_{M_i}$, that compose the dataset (the TOV solution for each EoS).\\

For the learning procedure, we have randomly split our dataset ($13038$ EoS) into two sets: training (80\%) and test (20\%).
The DNN/SVM-R models are trained on the training set and, once we select the model with highest performance, we determine its accuracy on the test set (the test set is only used once). For training, we have employed a 5-fold validation\footnote{This approach consists in splitting the train data into five partitions of equal size; then for each partition $i$, the model is trained on the remaining 4 partitions and evaluated on partition $i$. The final model accuracy is the average of the 5 accuracy values obtained.}. All data were normalized. The evaluation metric used is the Mean Squared Error (MSE), $\text{MSE}=(1/N)\sum_{i}(\widehat{y}_i-y_i)^2$, where $\widehat{y}_i$ is the model's prediction, $y_i$ is the true value, and $N$ is the number of points (EoS).

The DNN were trained using the Keras library \cite{chollet2017kerasR} with the TensorFlow backend \cite{tensorflow2015-whitepaper}. For the optimization procedure, we employed the Adam algorithm \cite{kingma2014adam} with a batch size of $32$ with the MSE loss function.  We used grid search to find the best DNN structure, i.e., the number of activation layers, the number of units per layer, and the activation functions. The best SVM-R model was also selected using a grid search, with the radial basis kernel function.   

\section{Results}
After selecting the best SVM-R and DNN models by grid search, we determine their 
accuracy on the test set. This is the generalization capacity of the models, i.e., their predictability power on unseen data (data not used for training).
The results are given in Table \ref{tab4:models_acc}.

\begin{table}[!htb]
	\centering
	\begin{tabular}{ccccc}
		\hline
		& \multicolumn{2}{c}{RMSE} & \multicolumn{2}{c}{$(\text{RMSE}/\bar{y})\times 100\%$} \\
		$\widehat{y}$	& DNN & SVM-R & DNN & SVM-R \\
		\hline
		\hline
		$\Lambda_{1.0M_{\odot}}$& $16.646$ & $23.547$ & $0.59$  & $0.84$ \\
		$\Lambda_{1.4M_{\odot}}$& $1.932$ & $2.236$ & $0.46$ & $0.54$ \\
		$R_{1.0M_{\odot}}$ [km] & $0.007$ & $0.012$ & $0.06$ & $0.10$ \\
		$R_{1.4M_{\odot}}$ [km] & $0.006$ & $0.010$ & $0.05$ & $0.08$ \\
		\hline
	\end{tabular}
	\caption{Root Mean Square Error (RMSE), $\sqrt{(1/N)\sum_{i}^N(\widehat{y}_i-y_i)^2}$ , for the DNN and SVM-R models on the test set ($N$ is the total number of EoS in the test set). We also show the root mean squared relative error, $\text{RMSE}/\bar{y}$, where $\bar{y}=(1/N)\sum_i^N y_i$ is the mean value of the $y$ ({\it true} value), expressed as a percentage.}
	\label{tab4:models_acc}
\end{table}

DNN shows higher accuracies than SVM-R. To compare the models performance on different scale quantities, we show the root mean squared relative error (in percentage) in Table \ref{tab4:models_acc}. The predictability power remains almost constant for both $\Lambda_{M_i}$ and $R_{M_i}$. 
For the canonical neutron star $1.4M_{\odot}$, we have a prediction error below 0.1\%  for $R_{1.4M_{\odot}}$ and  1\% for $\Lambda_{1.4M_{\odot}}$ for both the DNN and SVM-R analysis. 
Therefore, from the EoS empirical parameters, we can infer both $\Lambda_{1.4M_{\odot}}$ and $R_{1.4M_{\odot}}$ with an average error of $\Delta \Lambda = 1.932$ and $\Delta R=0.006$ km using DNN, and $\Delta \Lambda = 2.236$ and $\Delta R=0.010$ km using SVM-R. 
\\

Having successfully learned the non-linear functions between the feature space (empirical parameters) and the target space (neutron stars observables), one can now explore the concept of feature importance. It consists in measuring the impact of each empirical parameter (feature) on the $\Lambda_{M_i}$ and $R_{M_i}$ predictions. The importance of a specific empirical parameter is measured by the increase in the DNN/SVM-R model's prediction error after randomly permuting (shuffling) its values in the dataset, while keeping the remaining empirical parameters unchanged. 
In other words, we are analyzing how the models's mean squared residuals (the accuracy of the non-linear maps) depend on each empirical parameter. 
If an empirical parameter is important for the non-linear mappings (Eqs.\,(\ref{lmi})-(\ref{rmi})), shuffling its values considerably increases the prediction error, i.e.,  the models' predictions strongly rely on this empirical parameter. If, on the other hand,  the models' prediction error remains almost unchanged after shuffling a specific empirical parameter, we can say that the parameter has a smaller impact on the non-linear map. 
In order to estimate the effect, we determine the RMSE increase ratio, $\text{RMSE}^{*}/\text{RMSE}$ (see Table \ref{tab5:feature_importance}), where the $\text{RMSE}$ values are in Table \ref{tab4:models_acc} and 
$\text{RMSE}^{*}$ is the models' root mean square error when a specific empirical parameter was randomly permuted. The prediction error is calculated on the test set.\\

The increase on the relative RMSE ($\%$)  for the DNN prediction after the shuffling is applied is listed in Table \ref{tab5:feature_importance}. The results for the SVM-R agree well with  DNN results, i.e., the relative importance of each empirical parameter is almost the same. These results show that both methods are learning the same non-linear dependencies. Contrarily to the usual correlation analysis, which is only sensitive to linear dependencies, we are accessing herein the impact of each isolated parameter on the non-linear map while controlling for the other parameters. We applied this procedure to 1.0, 1.4 and 1.9 $M_\odot$ NS. The consideration of a massive star allows us to judge our assumption that the $Q_{\text{sat}}$ and $Q_{\text{sym}}$ parameters essentially determine the massive star properties.  

The $K_{\text{sat}}$ is the most important parameter in predicting both the $R_{M}$ and $\Lambda_{M}$, since this is the parameter that, for all masses considered, has the strongest effect on the radius and the tidal deformability with the shuffling. The sensitivity to $Q_{\text{sat}}$ is only observed for the largest mass considered. Within the isovector parameters, $K_{\text{sym}}$ is the one that most affects properties of  the two lowest masses considered, both $R_{1.0M_{\odot}}$ and $R_{1.4M_{\odot}}$, while for $R_{1.9M_{\odot}}$ it is $Q_{\text{sym}}$. The same pattern is seen for $\Lambda_{M}$. In fact, just considering the isovector properties, the results show that $\Lambda_{1.0M_{\odot}}$ and $\Lambda_{1.4M_{\odot}}$ are sensitive to $L_{\text{sym}}$, $K_{\text{sym}}$, and in a smaller extension to $Q_{\text{sym}}$, in accordance with \cite{Zhang_2018}, whereas $E_{\text{sym}}$ shows only a small impact on the tidal deformability, in agreement with \cite{Raithel_2019}.
Furthermore, our results confirm that the GW170817 event provides information on $K_{\text{sat}}$ and $K_{\text{sym}}$, in accordance with  \cite{Malik:2018zcf,Raithel_2019}.\\

\begin{table}[!htb]
	\centering
	\begin{tabular}{ccccccc}
		\hline
		& $R_{1.0 M_{\odot}}$ & $R_{1.4 M_{\odot}}$ &$R_{1.9 M_{\odot}}$ &
		$\Lambda_{1.0M_{\odot}}$ & $\Lambda_{1.4M_{\odot}}$ & $\Lambda_{1.9M_{\odot}}$\\ 
		\hline
		\hline
		$K_{\text{sat}}$ &  1.272	&	1.866	&	4.047	&  9.100	&   14.102	& 32.024\\
		$Q_{\text{sat}}$ &  0.259	&	0.606   &	2.433	&  2.220	&	4.897	& 19.454\\
		$E_{\text{sym}}$ &  0.494	&	0.249	&	0.120	&  1.672	&	1.779	& 1.979\\
		$L_{\text{sym}}$ &  0.996	&	0.838   &	0.820   &  5.401    &	4.509   & 7.089\\
		$K_{\text{sym}}$ &  1.054	&	1.211	&	2.211	&  6.325	&   8.352   & 15.456\\
		$Q_{\text{sym}}$ &  0.644	&	1.153	&	3.535	&  4.450	&   7.512	& 21.790\\
		\hline
	\end{tabular}
	\caption{Impact of each empirical parameter on the accuracy decrease  of the DNN predictions, measured by the increase of the RMSE (in percentage) when the values of each empirical parameter are randomly shuffled. 
	The increase of the RMSE is calculated by  $(\text{RMSE}^{*}/\text{RMSE})\times100\%$, where $\text{RMSE}^{*}$ 
is determined after randomly permuting all the values of a given empirical parameter that characterizes the dataset (see text for details). }
	\label{tab5:feature_importance}
\end{table}

\begin{figure*}[!htb]
	\centering
	\includegraphics[scale=0.32]{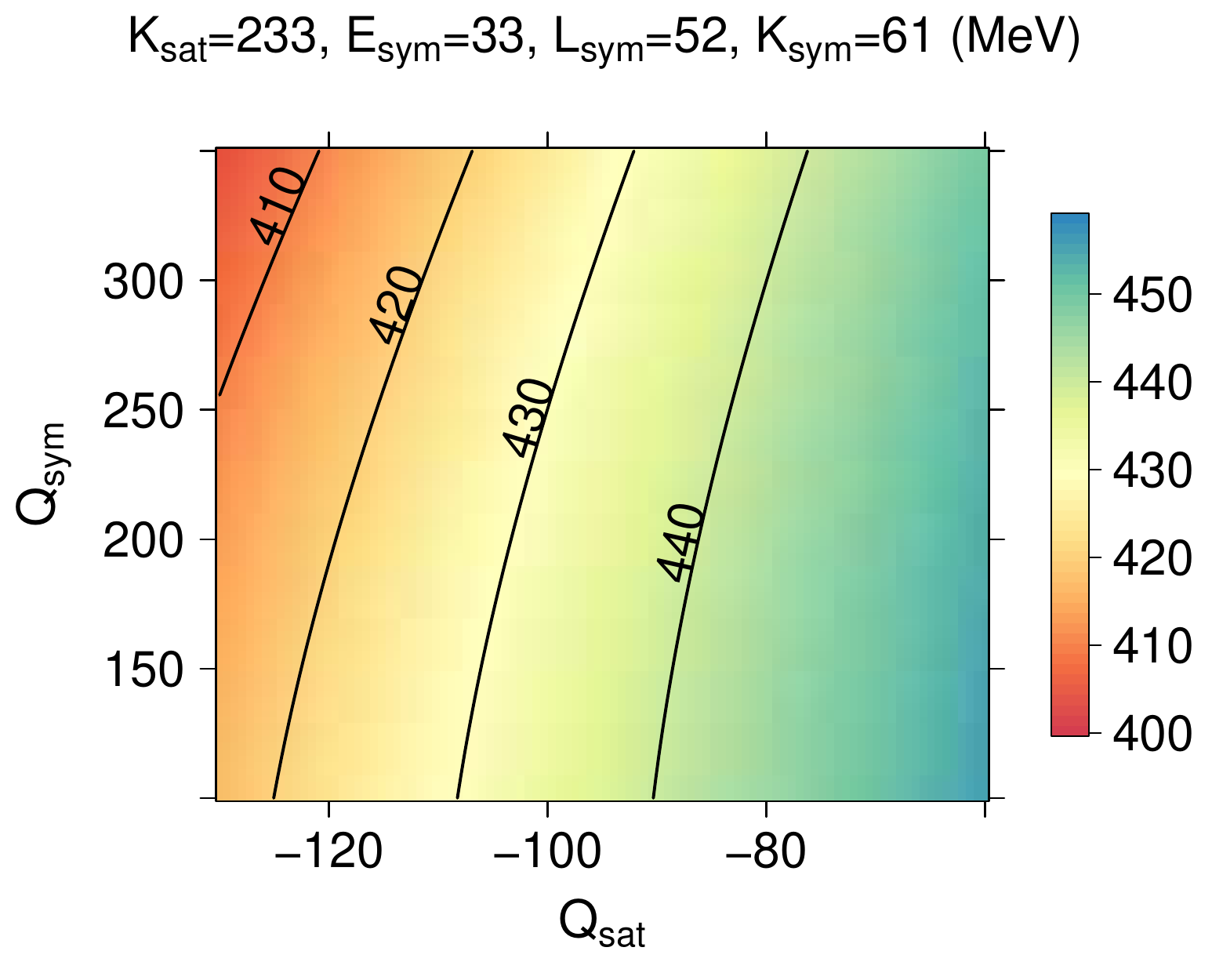}
	\includegraphics[scale=0.32]{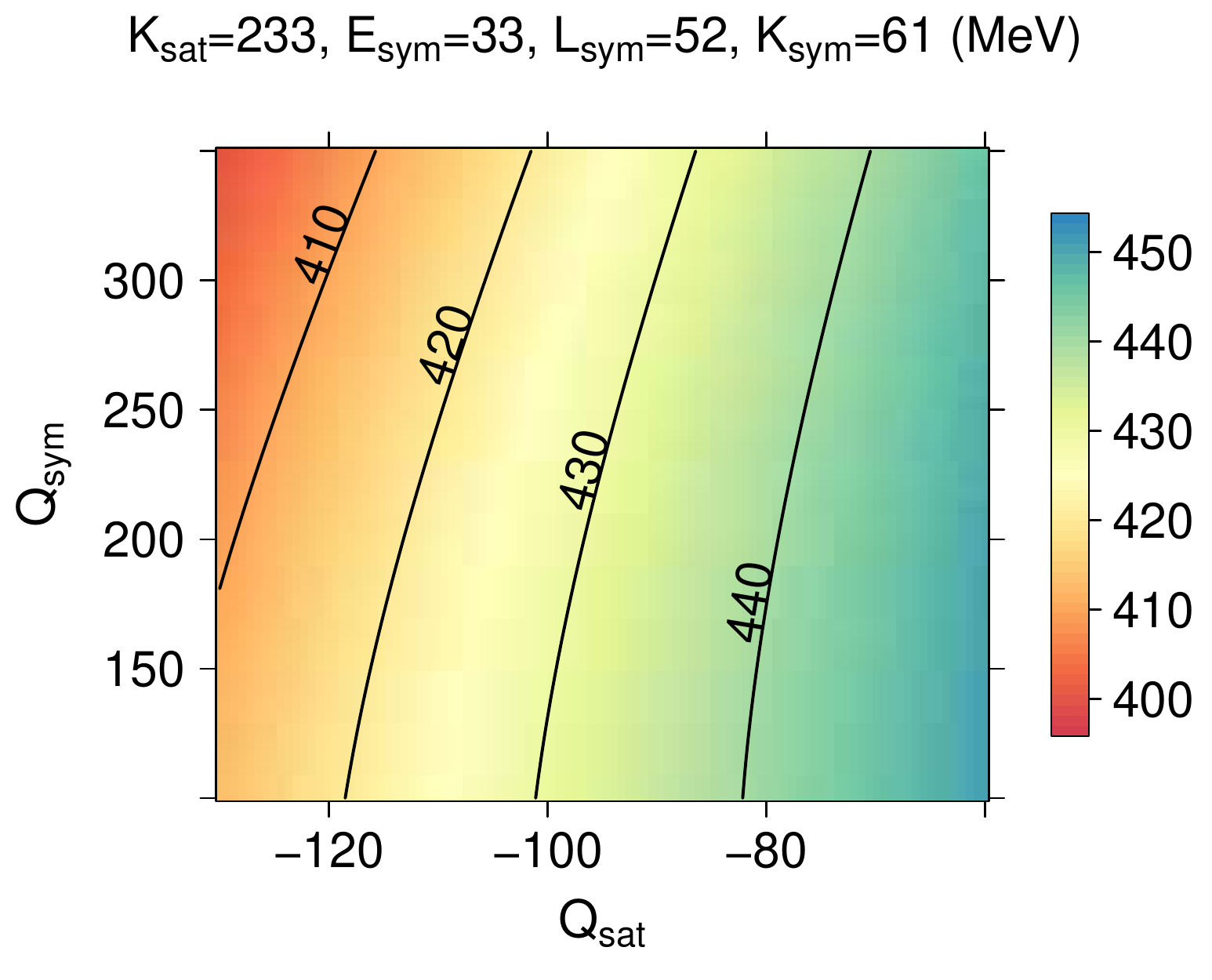}
	\includegraphics[scale=0.32]{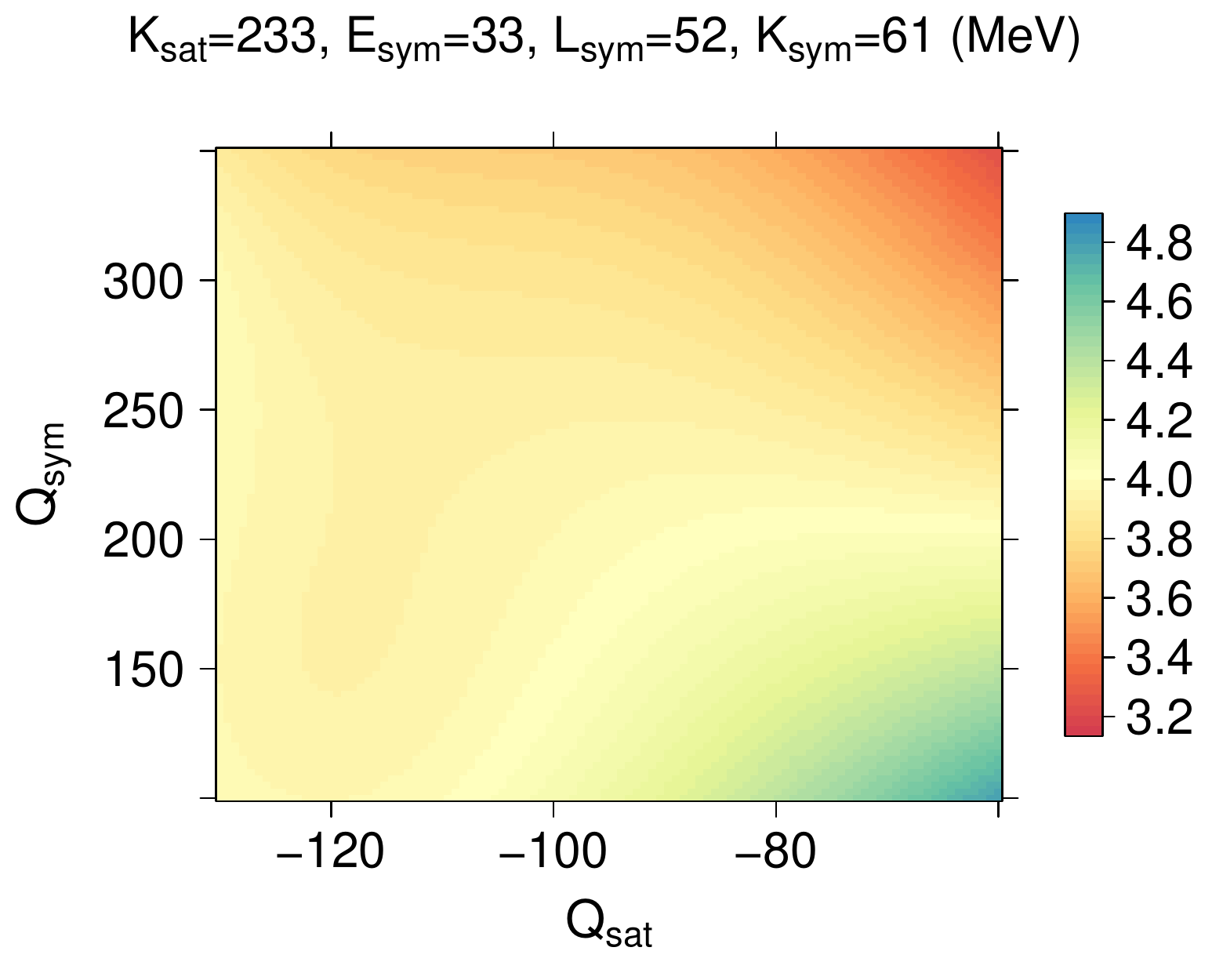}
	\caption{The DNN (left) and SVMR (center) predictions for $\widehat{\Lambda}_{1.4M_{\odot}}$ as a function of $Q_{\text{sym}}$ and $Q_{\text{sat}}$ (both in MeV) fixing: $K_{\text{sat}}=233$ MeV, $E_{\text{sym}}=33$ MeV, $L_{\text{sym}}=52$ MeV, and $K_{\text{sym}}=61$ MeV. The difference between DNN and SVMR predictions is shown in the right panel.}%
	\label{fig1:SMV_DNN}%
\end{figure*}

The relevance of having successfully learned the non-linear maps
between the empirical parameters and NS observables is that we are now
able to study their dependencies. Considering the big uncertainty on
the empirical parameters, we first analyze the impact of both
$Q_{\text{sym}}$ and $Q_{\text{sat}}$, which are two of the most uncertain empirical
parameters, on $\widehat{\Lambda}_{M_i}$ and $\widehat{R}_{M_i}$. To analyze $\widehat{\Lambda}_{M_i}(Q_{\text{sym}},Q_{\text{sat}})$ and
$\widehat{\Lambda}_{R_{i}}(Q_{\text{sym}},Q_{\text{sat}})$, we have fixed: $K_{\text{sat}}=233$ MeV, $E_{\text{sym}}=33$ MeV, and $L_{\text{sym}}=52$ MeV, close to the most probable values constrained by experiment. For  $K_{\text{sym}}$, we 
consider  $K_{\text{sym}}=\pm 61$ MeV taking into account our lack of
knowledge on this quantity. The symmetry energy curvature has been
constrained to the intervals $K_{\text{sym}}=-111.8 \pm 71.3$ MeV in \cite{Mondal2017}, $-112 \text{ MeV}<K_{\text{sym}}<-52 \text{ MeV}$  in \cite{Malik:2018zcf}, and $ -394 \text{ MeV}<K_{\text{sym}}<168 \text{ MeV}$  in \cite{Carson:2018xri}, and, recently, $-174 \text{ MeV}< K_{\text{sym}} <- 31 \text{ MeV}$  in \cite{Zimmerman2020}.
Figure \ref{fig1:SMV_DNN} shows the
$\widehat{\Lambda}_{1.4M_{\odot}}(Q_{\text{sym}},Q_{\text{sat}})$ map for DNN (left), SVM-R
(middle), and the difference between both predictions (right), taking  $K_{\text{sym}}=61$ MeV. Both  DNN and  SVM-R show an almost linear dependence,
i.e. $\widehat{\Lambda}_{1.4M_{\odot}}\sim aQ_{\text{sym}}+bQ_{\text{sat}}$. The same behavior is seen for $\widehat{R}_{1.4M_{\odot}}(Q_{\text{sym}},Q_{\text{sat}})$.
In fact, when we
subtract their predictions (right) we see that the $\widehat{\Lambda}_{1.4M_{\odot}}$ prediction discrepancy between both ML models is as low as $5$.

\begin{figure*}[!htb]
	\centering
	\includegraphics[scale=0.45]{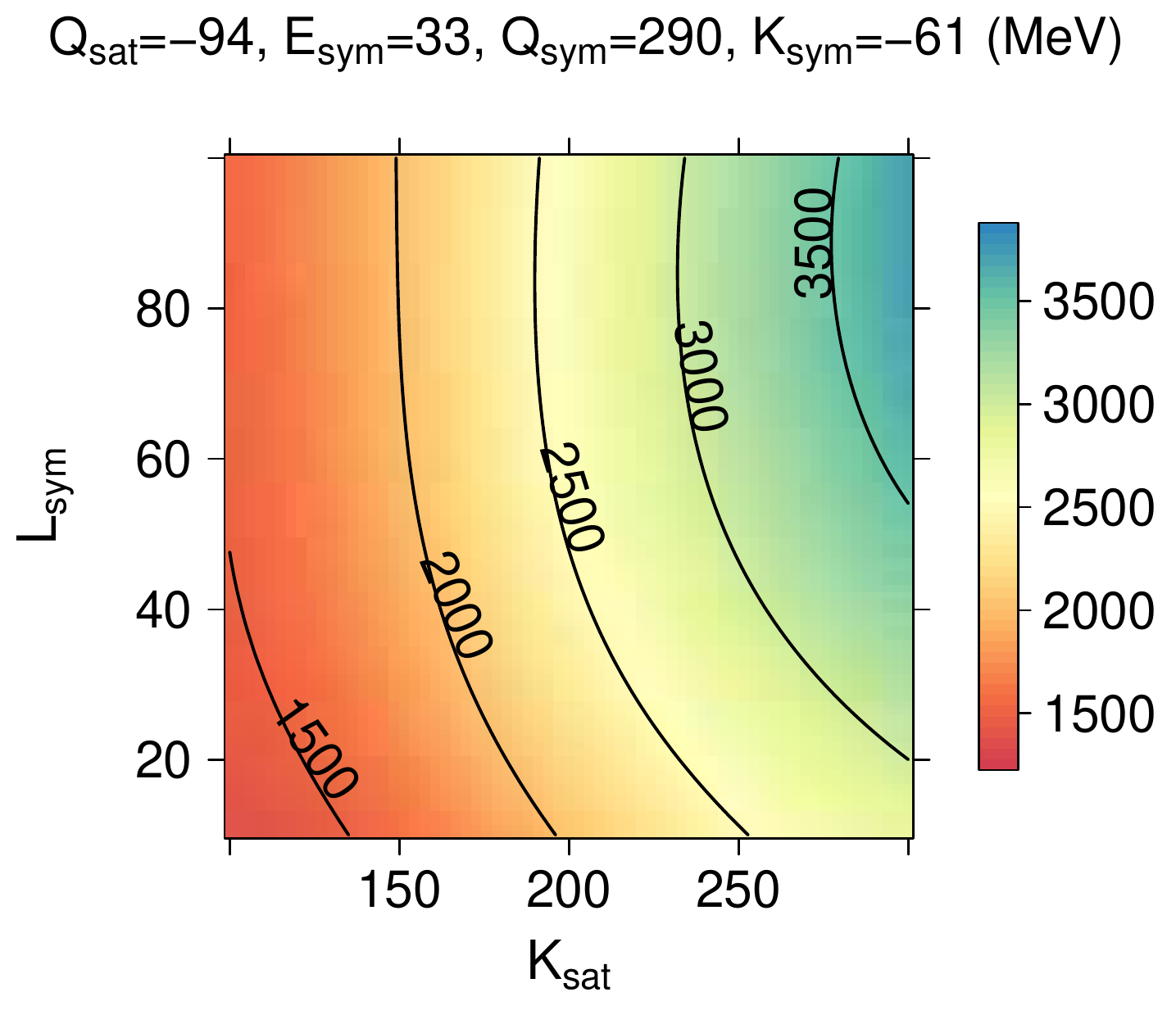}
	\includegraphics[scale=0.45]{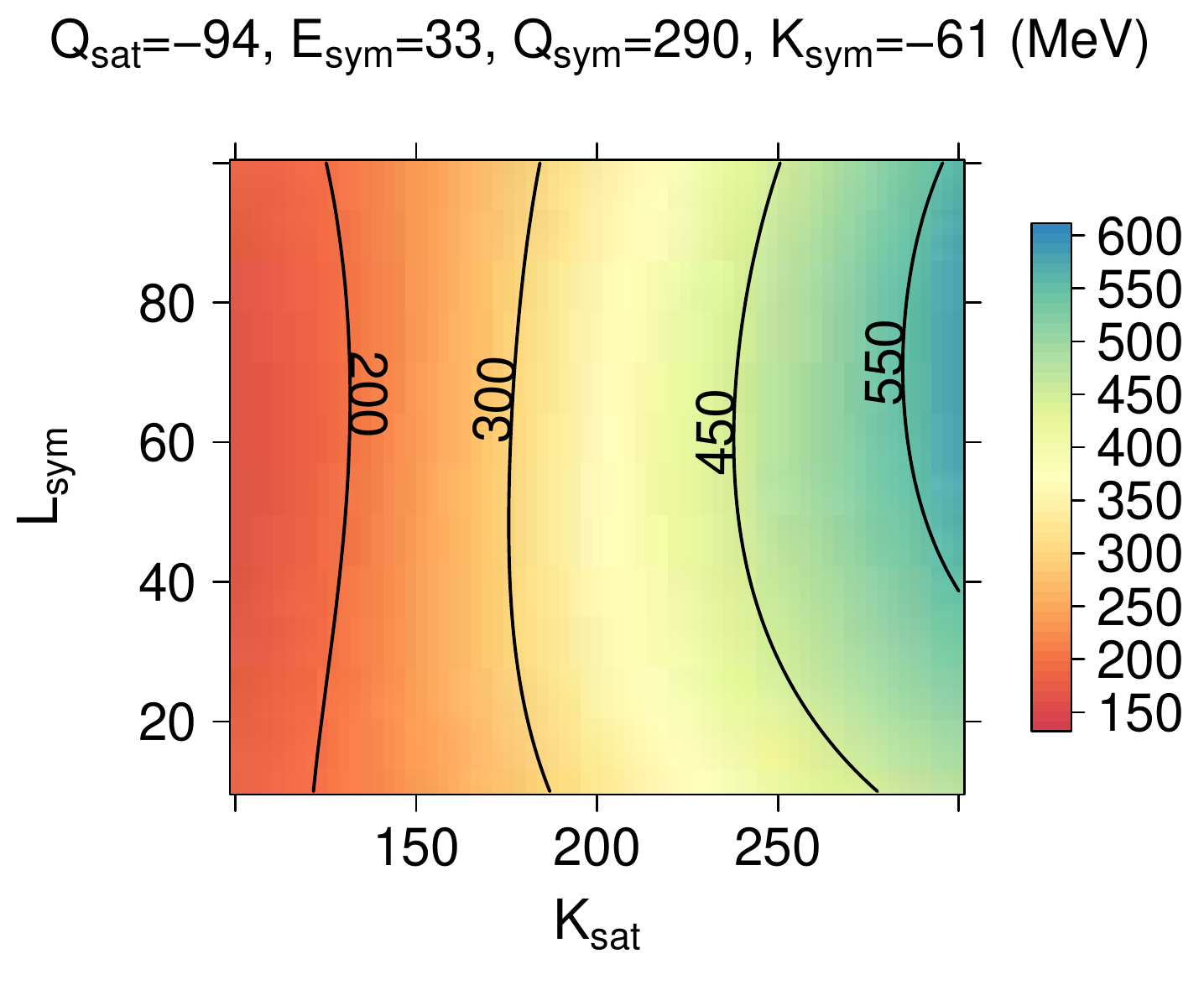}
	\caption{SVM-R predictions of $\widehat{\Lambda}_{M_i}(L_{\text{sym}},K_{\text{sat}})$ for $M_i=1.0M_{\odot}$ (left) and $M_i=1.4M_{\odot}$ (right), having fixed: $Q_{\text{sat}}=-94$ MeV, $E_{\text{sym}}=33$ MeV, $Q_{\text{sym}}=290$ MeV, and $K_{\text{sym}}=-61$ MeV.}%
	\label{fig3:first_set}%
      \end{figure*}

Hence, if the functions (\ref{lmi}) and (\ref{rmi}) are known, plots like the ones represented in Fig.	\ref{fig1:SMV_DNN} 
allow  us to visualize how the sensitivity obtained in Table \ref{tab5:feature_importance} depends on pairs of parameters, considering all the other parameters fixed. In the following, we will analyze two correlations studied in the literature, in particular, the  ($L_{\text{sym}}$, $K_{\text{sat}}$) correlation discussed in \cite{Alam2016,Malik:2018zcf,Carson:2018xri} and the  ($L_{\text{sym}}$, $K_{\text{sym}}$) discussed in \cite{Carson:2018xri,Carson:2019xxz}.

We now analyze the dependencies of $\widehat{\Lambda}_{M_i}$ and $\widehat{R}_{M_i}$ on
$L_{\text{sym}}$ and $K_{\text{sat}}$, i.e., the non-linear functions
$\widehat{\Lambda}_{M_i}(L_{\text{sym}},K_{\text{sat}})$ and $\widehat{R}_{M_i}(L_{\text{sym}},K_{\text{sat}})$, by
fixing  $Q_{\text{sat}}=-94$ MeV, $E_{\text{sym}}=33$ MeV, $Q_{\text{sym}}=290$ MeV, and $K_{\text{sym}}=-61$ MeV. 
The results for $\widehat{\Lambda}_{M_i}(L_{\text{sym}},K_{\text{sat}})$ and
$\widehat{R}_{M_i}(L_{\text{sym}},K_{\text{sat}})$ are plotted in Fig. \ref{fig3:first_set} and show a highly non-linear dependence. The first conclusion is the big effect of $K_{\text{sat}}$ on $\widehat{\Lambda}_{M_i}$, confirming the result in Table \ref{tab5:feature_importance} as the most important feature in explaining $\Lambda_{M_i}$. 
Furthermore, the low impact of $L_{\text{sym}}$ obtained from the feature importance study (see Table \ref{tab5:feature_importance}) is also seen 
in Fig. \ref{fig3:first_set}. 
The value of $\widehat{\Lambda}_{M_i}$ is almost insensitive to the value of
$L_{\text{sym}}$, although showing a slightly non-linear dependence. Increasing $K_{\text{sat}}$ makes $\widehat{\Lambda}_{M_i}$ and $\widehat{R}_{M_i}$ bigger. The $\widehat{R}_{M_i}(L_{\text{sym}},K_{\text{sat}})$ shows a trend similar to $\widehat{\Lambda}_{M_i}(L_{\text{sym}},K_{\text{sat}})$ including similar non-linear
dependencies. Let us stress that these results are just indicating
that after $\{Q_{\text{sat}},E_{\text{sym}},Q_{\text{sym}},K_{\text{sym}}\}$ have been fixed to
the above values, both $\widehat{\Lambda}_{M_i}$ and $\widehat{R}_{M_i}$ are sensitive to $K_{\text{sat}}$ and not to $L_{\text{sym}}$.

\begin{figure*}[!htb]
	\centering
	\includegraphics[scale=0.45]{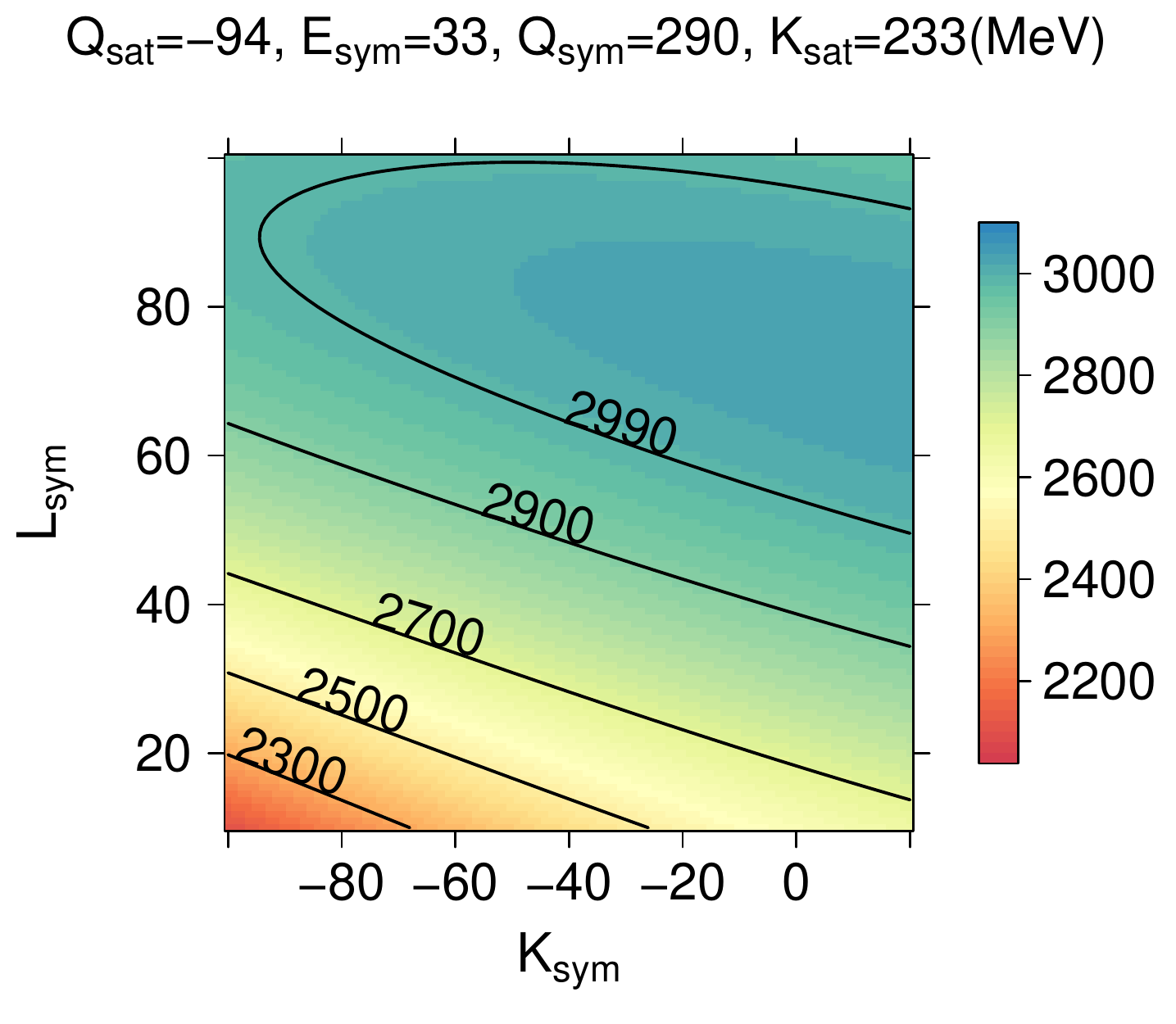}
	\includegraphics[scale=0.45]{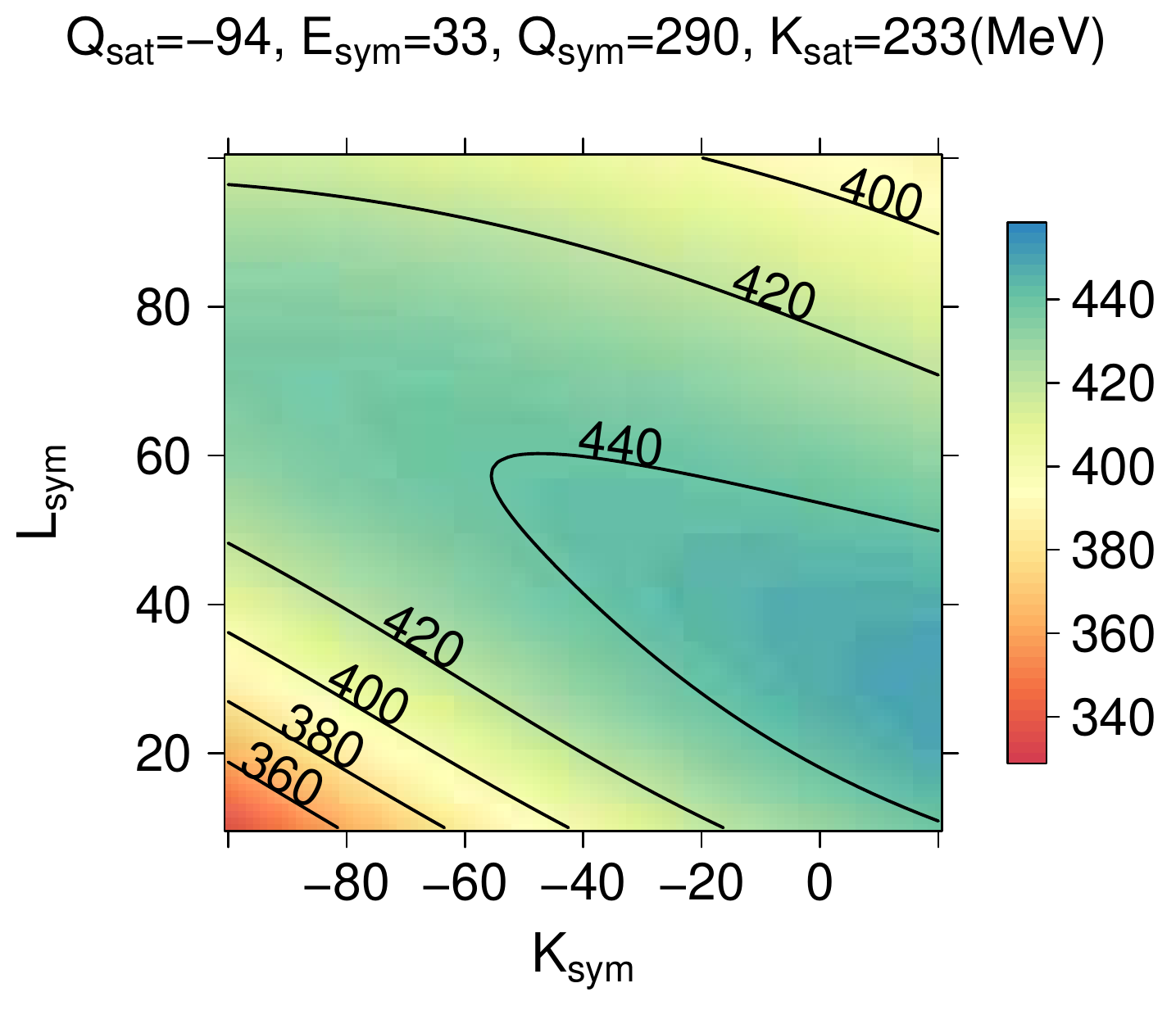}
	\caption{SVM-R predictions of $\widehat{\Lambda}_{M_i}(L_{\text{sym}},K_{\text{sym}})$ for $M_i=1.0M_{\odot}$ (left) and $M_i=1.4M_{\odot}$ (right), having fixed: $Q_{\text{sat}}=-94$ MeV, $E_\text{{sym}}=33$ MeV, $Q_{\text{sym}}=290$ MeV, and $K_{\text{sat}}=233$ MeV.}%
	\label{fig4:first_set}%
\end{figure*}

A strong correlation was found in \cite{Carson:2018xri} between the average tidal
deformability $\tilde\Lambda$, for all mass fractions compatible with
the GW170817, and the linear combination $L_{\text{sym}}+\lambda K_{\text{sym}}$.
In the following we analyze the dependencies of $\widehat{\Lambda}_{M_i}$ and
$\widehat{R}_{M_i}$ on $L_{\text{sym}}$ and $K_{\text{sym}}$, by fixing  $Q_{\text{sat}}=-94$ MeV, $E_{\text{sym}}=33$ MeV, $Q_{\text{sym}}=290$ MeV, and $K_{\text{sat}}=233$ MeV. The
results are shown in Fig. \ref{fig4:first_set}. It is interesting to notice
the existence of islands in the $(L_{\text{sym}},K_{\text{sym}})$ space that gives
the maximum values for $\widehat{\Lambda}_{M_i}$ and $\widehat{R}_{M_i}$. Fixing
$K_{\text{sym}}=-40$ MeV, for example, gives rise to a non-monotonic behavior
in $\widehat{\Lambda}(L_{\text{sym}})$ and $\widehat{R}(L_{\text{sym}})$. However, for low values of $L_{\text{sym}}$ and $K_{\text{sym}}$, there is an almost linear dependence on
$\widehat{\Lambda}_{M_i}$ and $\widehat{R}_{M_i}$. The importance of having the non-linear maps between NS observables and the empirical parameters is clear when
comparing Figures \ref{fig3:first_set} and \ref{fig4:first_set}. The effect of a parameter, e.g., $L_{\text{sym}}$, is only measurable once the set of remaining parameters are fixed and, furthermore, it crucially depends on their values.\\ 

A correlation between the NS radii and $K_{\text{sym}}$ 
is reported in \cite{Malik:2018zcf}, in particular if $M<1.4M_\odot$. Moreover, $K_{\text{sym}}$ and $K_{\text{sym}}+\gamma L_{\text{sym}}$ are seen to be strongly correlated with the tidal deformability for the star masses covered by GW170817 \cite{Carson:2018xri,Carson:2019xxz}. Figure \ref{fig4:first_set} reflects this fact: a linear dependence is found for $\widehat{\Lambda}_{1.0M_{\odot}}(L_{\text{sym}},K_{\text{sym}})$ and $\widehat{\Lambda}_{1.4M_{\odot}}(L_{\text{sym}},K_{\text{sym}})$ if one considers an EoS set characterized by $L_{\text{sym}}<60$ MeV and $K_{\text{sym}}<-20$ MeV. Furthermore, Fig. \ref{fig4:first_set} shows why the above correlations become weaker with increasing NS masses: their dependencies became essentially non-linear as $M$ increases. The NS radius shows similar dependencies.  
Universal relations between the NS observables
and the nuclear parameters captured by correlation analysis rely on the linear leading-order behavior of these non-linear maps. As soon as the non-linear dependencies become important, correlation analysis can no longer be used to constrain the nuclear parameters.

Let us point out an important feature seen in the left panel of Fig. \ref{fig4:first_set}.
It is possible to select subsets of points $\{L_{\text{sym}},K_{\text{sym}}\}$ (subset of EoS) that have either a positive or a negative correlation with $\widehat{\Lambda}$ (linear dependencies).  Furthermore, the correlations could vanish if a different subset  $\{L_{\text{sym}},K_{\text{sym}}\}$ is selected. This shows that discrepancies between different correlation analyses may arise from the kind of EoS used in each study.

\section{Conclusions}

In the present work we have shown how supervised ML  tools allow us to analyze NS properties, like the radius and tidal deformability, in terms of a set of six parameters that characterizes the EoS of stellar matter. Two methods have been tested, Deep Neural Networks (DNN) and Support Vector Machines Regression (SVM-R), and although similar, DNN shows higher accuracies.

First a set of more than $10^4$ EoS,  for which the pressure is an increasing function
of density and  the symmetry energy is non-negative, and describing stellar matter constrained by nuclear
matter properties,  observations and  causality, was generated.  ML methods were next applied using this set of models to learn the map between empirical parameters of the nuclear matter EoS and
astrophysical observables. 
These non-linear maps were then used to determine which nuclear matter parameters are the most important in describing NS properties as a function of its mass. This way we access how future GW events, depending on the NS masses involved, could be informative about the nuclear matter EoS, through their empirical parameters. From the analysis of the impact of each empirical parameter on the $\widehat{\Lambda}_{M_i}$ and $\widehat{R}_{M_i}$ predictions (Table \ref{tab5:feature_importance}),
we could conclude that $K_{\text{sat}}$ is the most important parameter in predicting both the $\widehat{R}_{M}$ and $\widehat{\Lambda}_{M}$, and $K_{\text{sym}}$ is the isovector parameter that most affects properties of stars with $M\lesssim 1.4 M_\odot$.
With the ML methods introduced, we could access the impact of each isolated parameter on the non-linear map while controlling for the other parameters. On the contrary, usual correlation analyses are  only sensitive to linear dependencies.

In order to test the properties of the non-linear map obtained between the
EoS parameters and the NS radius $\widehat{R}$ and tidal deformability $\widehat{\Lambda}$, we have studied in a two parameter plane $X_i,\, X_j$ how $\widehat{R}$ and $\widehat{\Lambda}$ behave, considering all the other parameters fixed. 
In particular, we have shown that the star radius and tidal deformability are  very sensitive to the incompressibility $K_{\text{sat}}$ and that, for  low mass stars ($M\lesssim 1.4 M_\odot$),  these properties are more sensitive to the  curvature $K_{\text{sym}}$
than to the slope $L_{\text{sym}}$. 
We have obtained a clear  linear dependence of
the tidal deformability on the linear combination $L_{\text{sym}}+\lambda
K_{\text{sym}}$ for the masses $M\lesssim 1.4 M_{\odot}$ if $K_{\text{sym}}$ is
small enough, otherwise a non-linear behavior is obtained which will
destroy a possible correlation between $L_{\text{sym}}+\lambda
K_{\text{sym}}$ and the tidal deformability. This drove us to the conclusion that  correlation analysis can not be used to constrain the nuclear parameters, if the non-linear dependencies become important.

A supervised ML  approach allows us to go beyond linear correlations and identify under which conditions astrophysical observations of neutron stars constrain the EoS of dense nuclear matter. In the future the EoS dataset used as  training set  will be generalized to a wider domain of acceptable models, including models that predict a first order phase transition. Let us, however, point out  that the present formalism, although allowing us to access information difficult to obtain otherwise,  should be considered as  complementary to other approaches. 
The main drawback is that the domain of validity of the ML maps is not automatically ensured, i.e, the ML maps are insensitive to possible "holes" on the  modeled hyper-surface due to the lack of training points in that region.

The present methodology might be useful when several (gravitational wave) astrophysical observations will be available. The non-linear maps, not being restricted to the linear dependencies, allow for more accurate inferences of the EOS parameter space from the knowledge of multiple NS observations.

Lastly, we have analyzed the impact of $M\geq 2.05M_\odot$ \cite{Cromartie2019} (68.3\% credibility interval) as an observational constraint, instead of $M\geq 1.97M_{\odot}$. Despite this stronger constraint not being realized by 2776 EoS, the results, using the set composed by the remaining valid EoS, are quite insensitive and all the conclusions remain the same.\\

\acknowledgments

This work was partly supported by 
Fundação  para  a Ciência e Tecnologia,  Portugal,
under the projects the projects UID/FIS/04564/2020 and POCI-01-0145-FEDER-029912 with financial support from POCI, in its FEDER component,
and by the FCT/MCTES budget through national funds  (OE),  
and by PHAROS COST Action CA16214.

\end{document}